\documentclass[aip,rsi,amsmath,amssymb,reprint,nofootinbib]{revtex4-2}

\usepackage{graphicx}
\usepackage{dcolumn}
\usepackage{bm}

\usepackage[utf8]{inputenc}
\usepackage[T1]{fontenc}
\usepackage{mathptmx}
\usepackage{etoolbox}

\usepackage{natbib}
\usepackage{gensymb}
\usepackage{booktabs}
\usepackage{xcolor}
\usepackage{colortbl}
\usepackage{comment}
\usepackage[hidelinks]{hyperref}
\graphicspath{ {Figures/} }

\begin{document}

\title{Deflected Beam Method for Absolute Current Density Determination} 

\author{R.H. Mattish}
\email[]{rickmattish@gmail.com}
\affiliation{Department of Physics and Astronomy, Clemson University, Clemson, SC 29634, USA}

\author{T.J. Burke}
\affiliation{Department of Physics and Astronomy, Clemson University, Clemson, SC 29634, USA}

\author{P.R. Johnson}
\affiliation{Department of Physics and Astronomy, Clemson University, Clemson, SC 29634, USA}

\author{C.E. Sosolik}
\email[]{sosolik@clemson.edu}
\affiliation{Department of Physics and Astronomy, Clemson University, Clemson, SC 29634, USA}

\author{J.P. Marler}
\email[]{jmarler@clemson.edu}
\affiliation{Department of Physics and Astronomy, Clemson University, Clemson, SC 29634, USA}

\date{\today}

\begin{abstract}
We present a broadly applicable in situ method for profiling ion beams using electrostatic deflectors and a Faraday cup. By deconvolving the detector geometry from the resulting current profiles, spatially resolved absolute current density profiles are obtained. We demonstrate this method's efficacy with low-density highly charged ion beams (specifically, Ne$^{8+}$). Details on experimental design are provided as well as the link to the deconvolution routine on Github.
\end{abstract}

\maketitle

\section{Introduction}
Determining the total flux of an ion beam is crucial for numerous applications; there are many ways to do so, the easiest perhaps being to focus the entire beam into a Faraday cup and measure the total current. For many studies, however, the determination of the spatially resolved absolute current density profile of ion beams is important in addition to total flux. This is true in particular for studies where only part of the ion beam will be overlapping with a crossed jet target or in surface-ion irradiation experiments where the interaction effects will be spatially resolved. Determining absolute charge exchange reaction rates between ions and a neutral gas, for example, necessitates knowing not only the neutral gas density but also the current density of the ion beam. While several techniques exist for determining the spatially resolved current density for certain beams, including surface damage distributions~\cite{assayag1993, wang1996}, infrared imaging,~\cite{davis1997} and phosphor viewing screens~\cite{ corr1992ion}, we introduce here a broadly applicable in situ method that is robust for beams with a range of fluxes and energies and can be implemented even in constrained vacuum conditions, eliminating some of the restrictions of other methods.

For example, it is possible to get a relative spatial profile of the current density via the method of observation of surface damage distributions. However, that requires physical examination of the sample and so either requires removing the sample from vacuum for inspection/preparation or an in-vacuum microscope and preparation station. On the other hand, infrared imaging of a surface requires an ion beam of sufficient energy (typically on the order of tens to hundreds of keV) to result in a measurable heating of the surface.~\cite{ yu2015} Phosphor screens have the additional cost of an external-to-vacuum sensor (e.g. a CCD camera) and must have a line of sight with a viewport. In addition, scintillating screens by themselves are only usable for high intensity ion beams. For low intensity beams, the addition of a microchannel plate is needed.~\cite{ pitters2019}

The method we describe here provides a quick in-vacuum determination of absolute current densities and can be used with both low and high energy ion beams with no additional hardware components outside of standard ion beam tools (i.e. electrostatic deflectors and a Faraday cup). This method is similar to the work of Ref.~\citenum{sosolik2000} in which a Faraday cup was mounted on a vacuum compatible translation stage. In addition to the inherent cost and space requirements associated with incorporating a translation stage, the data acquisition for a single profile took about 2 hours, leading to inherent errors from beam drift and limiting its usefulness for quick profile checks. Using electrostatic deflection of the beam across a fixed Faraday cup, we expect these data can be obtained in 10 minutes. The resolution of the spatial profile is still limited by the size of the Faraday cup.  Shown here is how this can be addressed by deconvolving the integrated current signal from the Faraday cup geometry.

In Section~\ref{Method}, we outline the method and provide details on the experimental components used to raster the beam and perform the measurements. We also provide an overview of the theory behind the deconvolution procedure used to remove the physical characteristics of the detector size from the measurements. In Section~\ref{Results}, we demonstrate this method using two different detector geometries and show the resulting spatially resolved ion beam current density profiles before and after a cathode-anode alignment to demonstrate one potential use case. In order for the method to be applicable to beams of different fluxes, careful consideration of the noise floor must be included in the deconvolution.

\section{Method}\label{Method}
The deflected beam method requires only a fixed Faraday cup (Fig.~\ref{FC3model}) mounted downstream from a set of x-y deflectors (Fig.~\ref{Totalsetup}). The measurement is made by rastering the beam across the cup face by scanning the x-y deflector voltages with a computer-controlled power supply, while an ammeter simultaneously records the current signal in the cup. The resulting two-dimensional integrated current map can be converted to a two-dimensional position-dependent current profile (Sec.~\ref{IonBeamRastering}) and then deconvolved from the detector geometry to obtain the absolute current density profile 
of the beam (Sec.~\ref{JDecon}).

\subsection{Faraday Cup Design}
\begin{figure}[t]
\includegraphics[width=\linewidth]{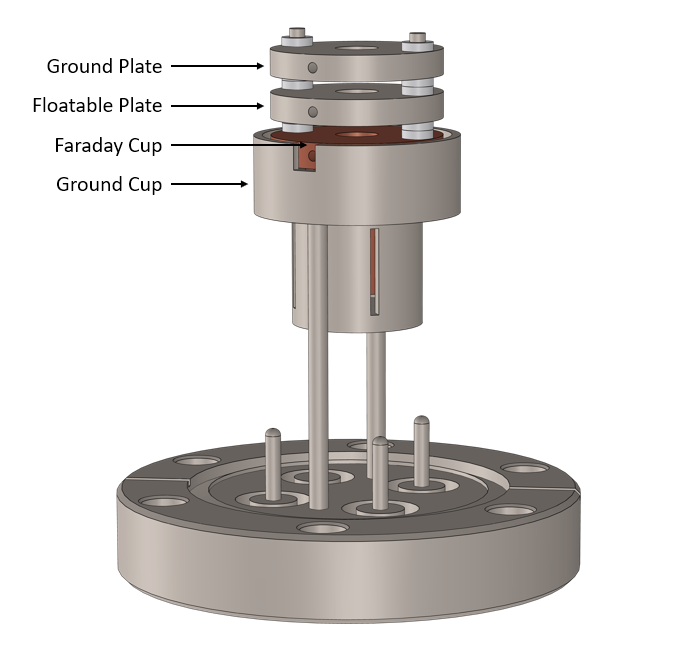}
\caption{Model of the Faraday cup assembly. The Faraday cup is mounted on a 2.75 in. ConFlat flange. From bottom to top, the assembly consists of a ground cup, a copper Faraday cup, a floatable plate, and a ground plate. The Faraday cup is further shielded by a sheet of grounded wire mesh (not shown) which encircles the assembly.\label{FC3model}}
\end{figure}

The Faraday cup (shown in Figure~\ref{FC3model}) used here is made out of OFHC copper and has an inner diameter of 6.35~mm. It is housed inside of a stainless steel cup which is grounded to shield the signal from stray charged particles. An electrically isolated stainless steel plate, 3.81~mm thick with a circular aperture of diameter 6.35~mm, sits 2.54~mm in front of the Faraday cup. This plate serves a dual purpose. For the current measurements, a negative voltage is applied for electron suppression. Prior to the measurements, this plate can be used as a retarding field analyser to determine the ion beam energy, which is necessary to calibrate the spatial deflection of the beam. 

Another stainless steel plate, 3.81~mm thick with a circular aperture of diameter 6.35~mm, sits 2.54~mm in front of the electrically-isolated plate. This faceplate is grounded to confine the electric field from the floatable plate. The faceplate hole also sets the size of the detection region. Therefore, the detector ``size'' can be varied cheaply with the same copper cup.  Here a second faceplate, identical to the first except with an aperture diameter of 1.00~mm, was also constructed.  

The entire stack was surrounded by a wire mesh on the sides to further screen stray electrons (e.g. those resulting from secondary emissions from ions hitting the vacuum chamber walls) or non-normal incidence ions. MACOR~\cite{Macor} tophats were used to provide electrical isolation.

\subsection{Ion Beam Rastering}\label{IonBeamRastering}
A Keithley 2000 multimeter (with a serial connection to the computer) was used in conjunction with a femtoammeter to measure the ion beam current collected in the Faraday cup. A computer-controlled power supply was used to set the voltages on the x and y deflectors. A custom Python script was then used to vary and record the x and y deflector voltages and record the ion beam current collected in the Faraday cup. While the exact acquisition time will vary depending on the specific hardware used for rastering, the raster area size, and the step size, we expect that profiles such as those we present here can be obtained in about 10 minutes.

The spatial deflection of the ion beam was calculated from the axial velocity of the ions, the radial acceleration from the deflector potentials, and the known geometry of the deflectors and distance between the deflectors and the Faraday cup. Here we use the following directional notation: axis of beam propagation ($z$) and horizontal ($x$) and vertical ($y$) components of the radial direction. The axial velocity of the ions, $v_{z}$, is determined by the ion source potential, $V_{s}$, used to accelerate the ions and is given by:
\begin{equation}\label{velocity_equation}
v_{z} = \sqrt {\frac{2qV_{s}}{m}}
\end{equation}
where $q$ and $m$ are the charge and mass of the species of ion. If $V_s$ is not known, it can be measured using a retarding field analyser. Using the parallel-plate approximation, the radial acceleration from the deflectors is:
\begin{equation}\label{acceleration_equation}
a = \frac{q}{m} \frac{\Delta V_d}{d}
\end{equation}
where $\Delta V_d$ and $d$ are the potential difference and separation distance of the parallel deflector plates, respectively. Using these to solve the kinematics equations, we obtain an equation for converting from applied deflector voltage to  deflection distance at the location of the detector which is charge and mass invariant:
\begin{equation}\label{deflection_equation}
\Delta y = \frac{\Delta V_d}{2dV_{s}}\left(\frac{\Delta z_1^2}{2}+\Delta z_1\Delta z_2\right)
\end{equation}
where $\Delta y$ is the amount of off-axis deflection of the beam center induced by the deflectors, $\Delta z_1$ is the path length of the uniform electric field region within the plates (i.e. the length of the deflector plates along the axis of propagation), and $\Delta z_2$ is the length of the field-free region between the deflector and the Faraday cup. 

\begin{figure*}[t]
\includegraphics[width=\textwidth]{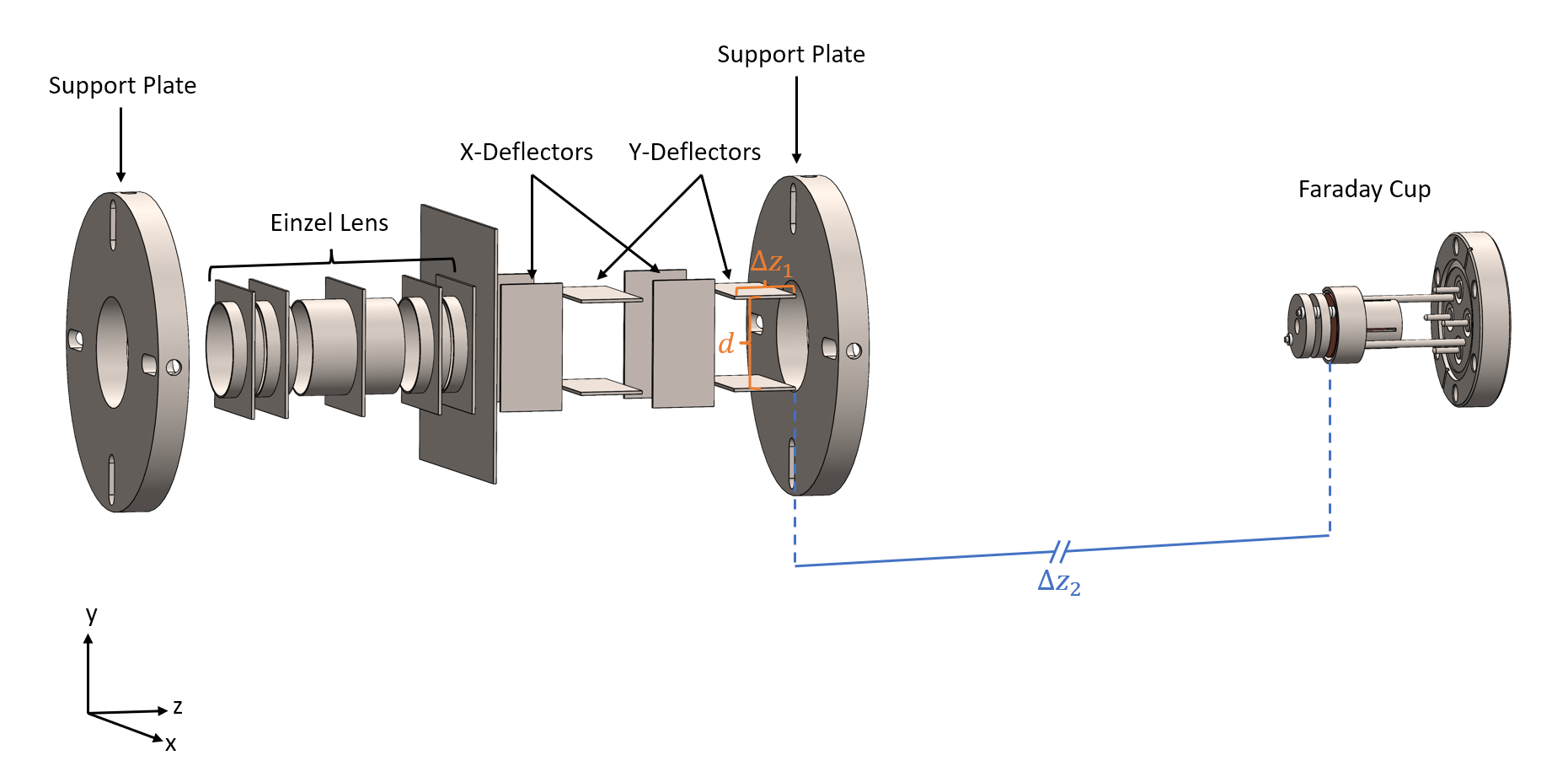}
\caption{Model of the in-vacuum experimental apparatus for ion beam characterization. From left to right: support plate, einzel lens, two pairs of x and y deflectors, support plate, Faraday cup.\label{Totalsetup}}
\end{figure*}

A CAD model of the ion optics that were used for the ion beam rastering is shown in Figure~\ref{Totalsetup}. The side length of a deflector plate along the axial direction, $\Delta z_1$, is 2.54~cm, and the separation distance $d$ between the deflector plates is 3.81~cm. The length of the field-free region between the last pair of deflectors and the Faraday cup, $\Delta z_2$, is 48~cm. For the measurements presented in this paper, the source potential used to accelerate the ions, $V_s$, was 4000 V.  For Ne$^{8+}$, this resulted in an ion beam energy of $q\cdot V_s=32$ keV. To give a sense of the deflection resulting from the described apparatus, a potential difference of 50~V (applied as $\pm 25$~V) across a pair of deflector plates resulted in 2~mm of off-axis deflection at the Faraday cup.

\subsection{Deconvolution Procedure}\label{JDecon}
To obtain the absolute current density profile of the ion beam, we must first deconvolve it from the geometry of the detector . Here we follow the procedure of Sosolik et al.~\cite{sosolik2000} which is discussed at length in Appendix A of Ref.~\citenum{dahl1998}. A concise summary of the method is given here.

The measured integrated current profile, $i(\mathbf{r})$, is a convolution of the current density profile, $j(\mathbf{r})$, and the detector function of the faceplate aperture, $d(\mathbf{r})$. This is expressed mathematically as:
\begin{equation}\label{convolution_equation}
i( \mathbf{r} )=\int d( \mathbf{r} - \mathbf{r} ') j( \mathbf{r} ') d \mathbf{r} '.
\end{equation}

From the convolution theorem~\cite{Arfken2001, bracewell1965}, we can take the Fourier transform of Equation~\ref{convolution_equation} to reduce the integral to a simple matrix multiplication:
\begin{equation}\label{FT_convolution_equation}
I( \mathbf{k} )=D( \mathbf{k} ) J( \mathbf{k} )
\end{equation}
where $I( \mathbf{k} )$, $D( \mathbf{k} )$, and $J( \mathbf{k} )$ are the Fourier transforms of $i( \mathbf{r} )$, $d(\mathbf{r})$, and $j(\mathbf{r})$, respectively. 

To obtain $I( \mathbf{k} )$, we simply take the discrete Fourier transform of the $i( \mathbf{r} )$ data arranged in matrix form. For a circular aperture detector of radius $R_0$, $d(\mathbf{r})$ can be represented as:
\begin{equation}\label{detector_function}
d(\mathbf{r})=
	\begin{cases} 
      1 & \text{if } r \leq R_0\\
      0 & \text{if } r > R_0
   \end{cases}
\end{equation}
Taking the Fourier transform of Equation~\ref{detector_function}, we obtain:
\begin{equation}\label{FT_detector_function}
D( \mathbf{k} )=\frac{2 \pi R_0}{k} J_1 (kR_0)
\end{equation}
where $J_1$ is a Bessel function of the first kind and $k=|\mathbf{k}|$.

Having obtained both $I( \mathbf{k} )$ and $D( \mathbf{k} )$, it is tempting to simply solve for $J( \mathbf{k} )$ using Equation~\ref{FT_convolution_equation} via division. Aside from the issue that $D( \mathbf{k} )$ may contain zeros, that approach fails to take into account noise in the signal which, if not accounted for, will be magnified wherever $D( \mathbf{k} )$ is small. Instead, Equation~\ref{FT_convolution_equation} can be rewritten as:
\begin{equation}\label{FT_current_density}
J( \mathbf{k} )=\frac{I( \mathbf{k} )D( \mathbf{k} )^*}{|D( \mathbf{k} )|^2+\eta (\mathbf{k})}
\end{equation}
where $\eta (\mathbf{k})$ is defined as the Fourier transformed noise to current density power ratio:
\begin{equation}\label{Eta_definition}
\eta (\mathbf{k})\equiv \frac{|N( \mathbf{k} )|^2}{|J( \mathbf{k} )|^2}
\end{equation}
If there is no noise, $\eta (\mathbf{k})=0$ and Equation~\ref{FT_current_density} reduces to Equation~\ref{FT_convolution_equation}. Taking the inverse discrete Fourier transform of Equation~\ref{FT_current_density} yields an estimate of the absolute current density $j(\mathbf{r})$.

This deconvolution procedure is algorithmic save the determination of $\eta (\mathbf{k})$, which must be performed for each experimental system. By definition, $\eta (\mathbf{k})$ cannot be determined a priori since it depends on $J( \mathbf{k} )$. In practice, however, it can be estimated reasonably well from the azimuthally averaged power spectrum of the Fourier transformed current signal, $\langle |I( \mathbf{k} )|^2\rangle$. More details on estimating $\eta (\mathbf{k})$ are given in Section~\ref{eta}.

\section{Results}\label{Results}
\subsection{Current Density}
\begin{figure*}[t]
\includegraphics[width=\textwidth]{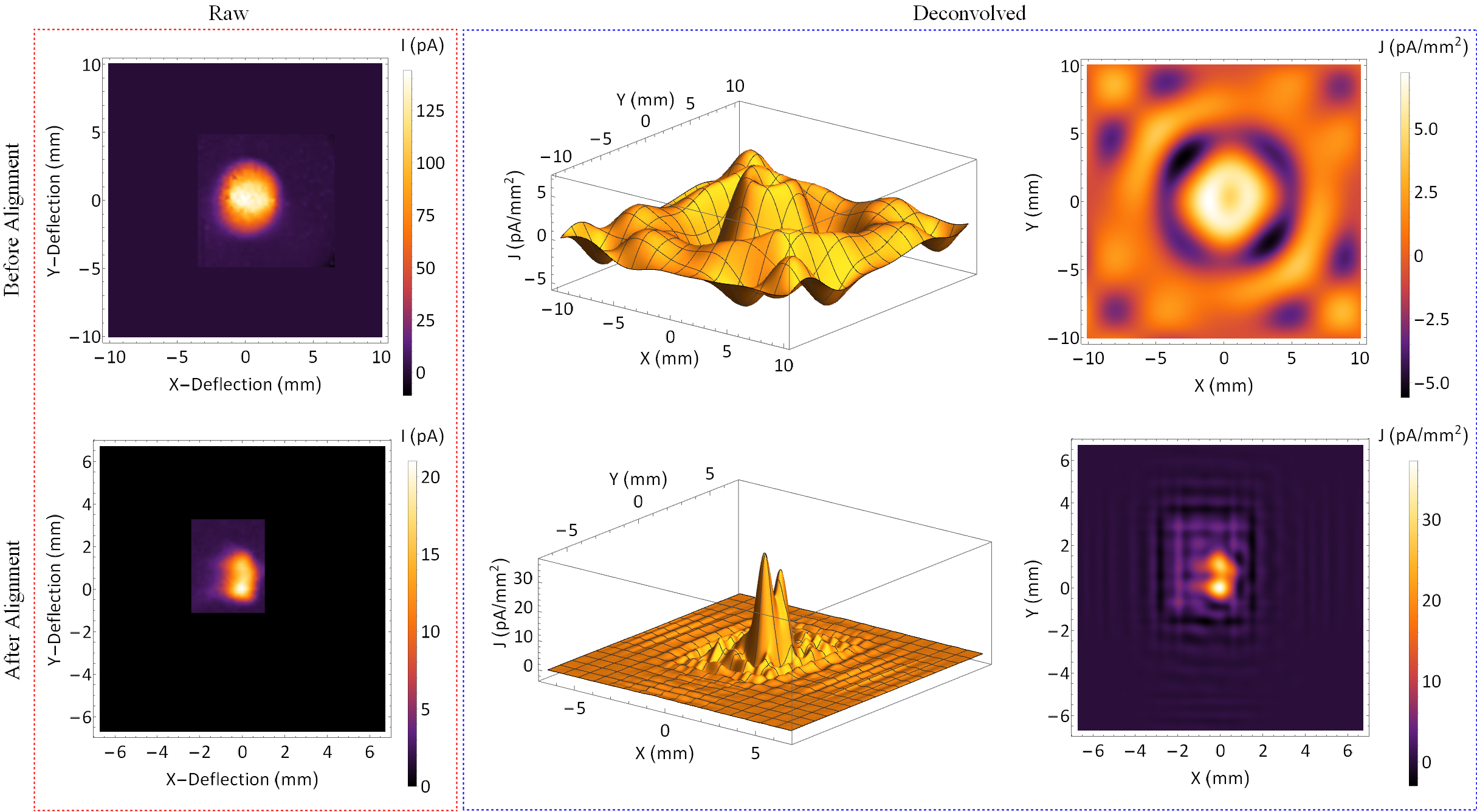}
\caption{Combined plots showing before and after deconvolution for ion beams obtained before (top row) and after (bottom row) cathode-anode alignment. Raw current measurements as a function of ion beam deflection are shown in the box on the left. Deconvolved absolute current density measurements as a function of spatial position are shown as both 3D and 2D plots inside the box on the right.\label{JComparison}}
\end{figure*}
To demonstrate this method, the Clemson University elec-
tron beam ion trap (CUEBIT) was used to generate highly
charged ions (HCIs)~\cite{shyam2015}. Charge-state selected ions were extracted and  passed through an einzel lens and two sets of x and y electrostatic deflectors (as shown in Figure~\ref{Totalsetup}). Only the closest pair of deflectors in each direction was used to raster the beam across the Faraday cup. 

$\Delta z_2$ is different for the x and y deflectors, which results in a slight difference in the amount of deflection in the two directions given the same deflector voltage. Fourier transforms require uniformly spaced data points, however. We decided that the easiest way to satisfy this criteria was to interpolate the data over a uniformly spaced grid. Alternatively, deflectors could have been constructed so as to be equidistant from the Faraday cup.

Here, a beam of Ne$^{8+}$ ions with a total beam current of about 140~pA  was measured when focused in the cup using the large (6.35~mm) aperture faceplate. The beam was then rastered across the Faraday cup using the electrostatic deflectors, and the measured current data was deconvolved. An alignment of the cathode was performed with respect to the anode in the CUEBIT source, and the experiment was then repeated with the small (1~mm) aperture faceplate with a higher intensity beam to see how the alignment affected the beam geometry.

The results are shown in Figure~\ref{JComparison}. The beam profiled before the alignment had a maximum current density of 7.18~$\pm$~0.16~pA/mm$^2$ and exhibited a ring shape. The beam profiled after the alignment had a maximum current density of 37.2~$\pm$~2.5~pA/mm$^2$ and exhibited a bimodal structure.

It is worth noting that the deconvolved current density plots have some circular ripples in them which are not present in the raw data. These spherical waves are an artifact of the deconvolution procedure and result from the detector function in Fourier space for a circular aperture detector (Equation~\ref{FT_detector_function}). While we have left these ripples in the data for completeness, in practice they can be suppressed by cleaning the $i(\mathbf{r})$ data matrix. This can be done by multiplying by a function which is one everywhere there is signal and zero everywhere there is only background noise with a sharp but continuous transition between these two values.

The change in beam geometry from a quasi-circular beam to a bimodal beam reflects a change in the CUEBIT source dynamics. The cathode-anode alignment that was performed between the measurements should have altered the electron beam space charge in the source region. This demonstrates a useful application of this method; it enables the study of trap physics.

\subsection{Determining $\eta (\mathbf{k})$}\label{eta}
\begin{figure}[htbp!]
\includegraphics[width=\linewidth]{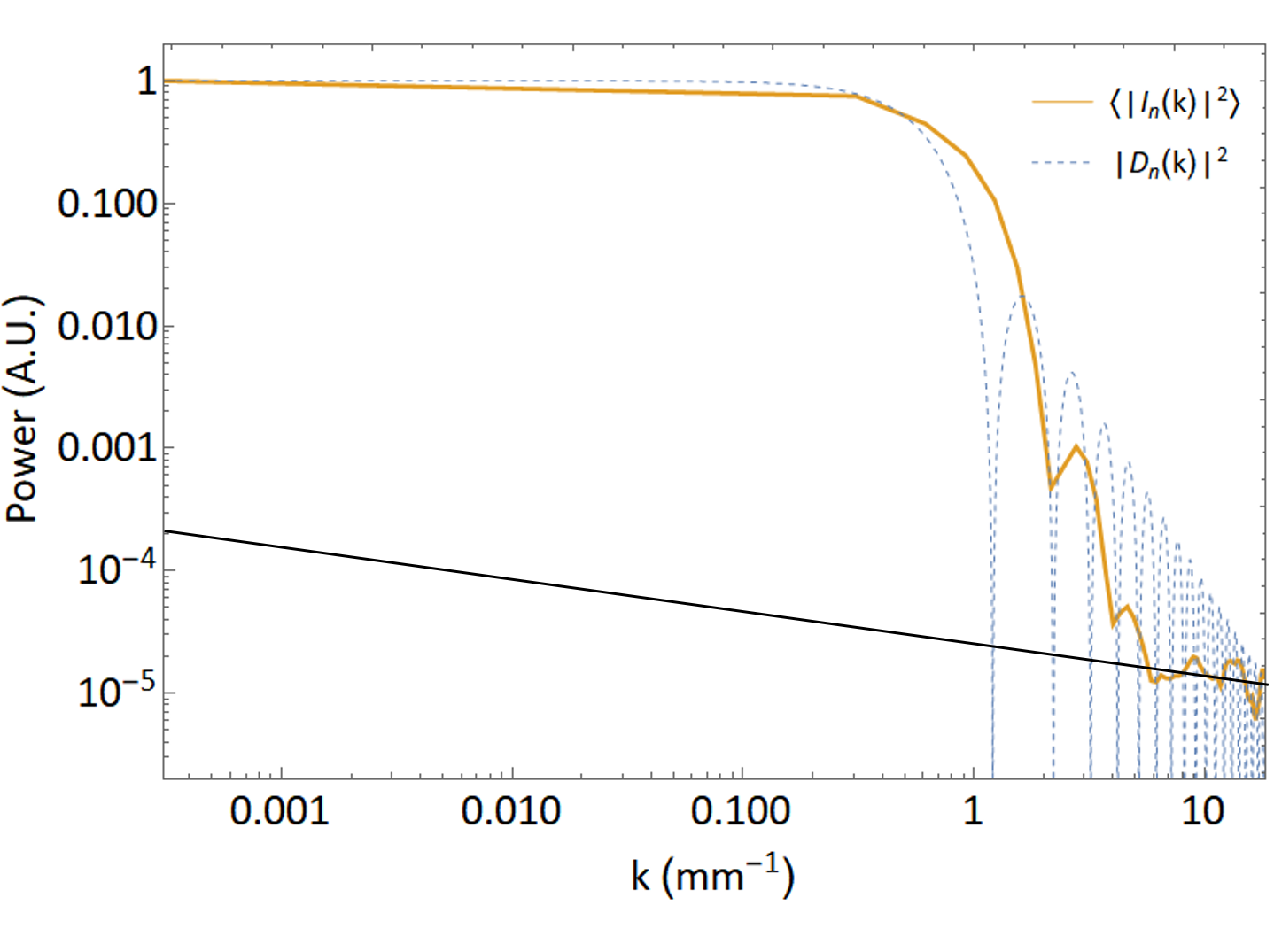}
\caption{The azimuthally averaged, normalized power spectrum of $I (\mathbf{k})$ is plotted as the solid orange line, and the normalized power spectrum of $D (\mathbf{k})$ is plotted behind it for reference as the dashed blue line. The black line provides a rough estimate of the noise.\label{DeterminingEta}}
\end{figure}
As mentioned in Section~\ref{JDecon}, it is necessary to obtain $\eta (\mathbf{k})$ to use as an input for Equation~\ref{FT_current_density}.  Here, we give one method which can be used for estimating $\eta (\mathbf{k})$.

In this estimate, it was assumed that the noise was independent of the signal strength; a reasonable assumption since the multimeter generally contributes additive noise. The measured (noisy) signal can be written as:
\begin{equation}\label{noisy_signal}
I( \mathbf{k} )=D( \mathbf{k} ) J( \mathbf{k} ) + N( \mathbf{k} )
\end{equation}
Additionally, it was assumed that $N( \mathbf{k})$ is a constant, independent of $\mathbf{k}$. With this assumption,  the $\mathbf{k}$-dependence of $\eta (\mathbf{k})$ comes only from $J (\mathbf{k})$. An estimate of $J (\mathbf{k})$ was obtained by using a constant $\eta =0.01$ in Equation~\ref{FT_current_density}. The resulting matrix was then used to calculate a normalized current density power spectrum, which we denote with a subscript n as $|J_n( \mathbf{k} )|^2$. It was normalized such that $|J_n(0)|^2=1$. This produced an estimate of the dependence of $\eta ( \mathbf{k})$ on $\mathbf{k}$ without regard to the magnitude of  $\eta$.

According to Equation~\ref{noisy_signal}, $I( \mathbf{k} )\approx J( \mathbf{k} )$ when $D( \mathbf{k} )$ is large (i.e. when $D( \mathbf{k} )\simeq 1$), which occurs at small $\mathbf{k}$. Likewise, $I( \mathbf{k} )\approx N( \mathbf{k} )$ when $D( \mathbf{k} )$ is small (i.e. when $D( \mathbf{k} )\simeq 0$), which occurs at large $\mathbf{k}$. In summary, at small $\mathbf{k}$ the signal predominates and at large $\mathbf{k}$ the noise predominates. Therefore, a reasonable estimate of the magnitude of $\eta (\mathbf{k})$ can be made by looking at the power spectrum of $I (\mathbf{k})$.

The azimuthally-averaged, normalized power spectrum of the current profile, $\langle |I_n( \mathbf{k})|^2\rangle$, obtained using the larger aperture faceplate is plotted in Figure~\ref{DeterminingEta} along with the normalized power spectrum of the detector function, $|D_n( \mathbf{k})|^2$, for reference. Both of these spectra were normalized at $\mathbf{k}=0$. The straight black line represents a rough power law fit to the part of the power spectrum that is assumed to be predominantly noise. By doing so, the noise level was extrapolated back to $\mathbf{k}=0$, and the noise-to-signal power ratio was estimated. This magnitude estimate, $\eta(0)$, can be combined with the estimate of the $\mathbf{k}$-dependence, $|J_n( \mathbf{k} )|^2$, to obtain an estimate of $\eta ( \mathbf{k})$:
\begin{equation}\label{Eta_estimate}
\eta (\mathbf{k})\approx \frac{\eta(0)}{|J_n( \mathbf{k} )|^2}
\end{equation}

In the case of the power spectrum shown in Figure~\ref{DeterminingEta}, the magnitude of the noise-to-signal power ratio at $\mathbf{k}=0$ was estimated to be $\eta(0)=2\times 10^{-4}$. This magnitude was used to obtain an estimate of $\eta ( \mathbf{k})$ from Equation~\ref{Eta_estimate}, which was subsequently used to solve Equation~\ref{FT_current_density} for $J( \mathbf{k})$ and inverse Fourier transformed to obtain the $j( \mathbf{r})$ shown in Figure~\ref{JComparison}.

\begin{figure}[t!]
\includegraphics[width=\linewidth]{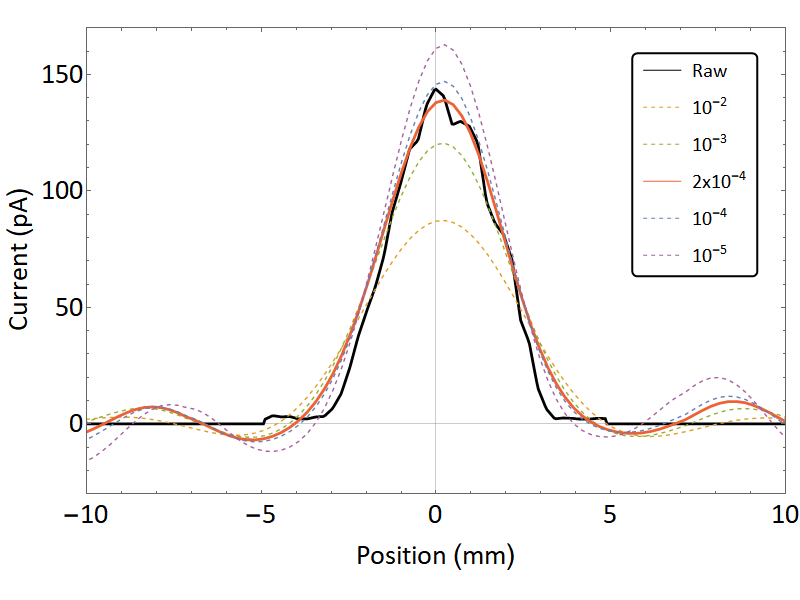}
\caption{Comparison of the raw current measurements (solid black) to current obtained using our estimate of $\eta(0)=2\times 10^{-4}$ (solid red-orange) along with those obtained using other various values for $\eta(0)$.\label{ComparingEta}}
\end{figure}

To test the accuracy of this estimate, we have recalculated the expected integrated current using Equation~\ref{convolution_equation} and the current density matrices obtained for a range of values for $\eta(0)$ and compared these expected current profiles to the measured current data. The resulting current profiles are plotted in Figure~\ref{ComparingEta} along with the raw measured current profile. Our estimate of $\eta(0)=2\times 10^{-4}$ produces the closest match.

\section{Conclusion}
We have presented an in situ method for obtaining the spatially resolved absolute current density profile of ion beams using low-cost, standard type components  (i.e. Faraday cups and electrostatic deflectors). This method is applicable for a broad range of energies and can be implemented in constrained vacuum environments. We have demonstrated its use with two different detector geometries (a 6.35~mm and a 1.00~mm diameter circular aperture) using ion beams generated at the CUEBIT. The results showed previously unknown structure in the beams and revealed the effect of the cathode-anode alignment on the extracted beam profiles.

The use of electrostatic deflectors means the implementation is cheaper and easier to  install than a two-axis translation stage. In addition, they can be readily computer-controlled which speeds up the rate at which a raster-scan can be performed, leading to an order of magnitude reduction in the acquisition time. The deconvolution procedure has been turned into a computer algorithm, and the Python script is available on Github for download and use~\cite{deconvolution_script}. Instructions for use as well as sample input files are available in the same Github repository.

\begin{acknowledgments}
The authors would like to thank the Physics and Astronomy Instrumentation Shop for machining the parts for the Faraday cup used for the ion beam current measurements. This work was supported in part by NASA grant APRA-80NSSC19K0679 and NSF grant 1815932.
\end{acknowledgments}

\section*{Data Availability}
The data that support the findings of this study are available from the corresponding author upon reasonable request.

\bibliography{Sources}

\begin{thebibliography}{13}%
\makeatletter
\providecommand \@ifxundefined [1]{%
 \@ifx{#1\undefined}
}%
\providecommand \@ifnum [1]{%
 \ifnum #1\expandafter \@firstoftwo
 \else \expandafter \@secondoftwo
 \fi
}%
\providecommand \@ifx [1]{%
 \ifx #1\expandafter \@firstoftwo
 \else \expandafter \@secondoftwo
 \fi
}%
\providecommand \natexlab [1]{#1}%
\providecommand \enquote  [1]{``#1''}%
\providecommand \bibnamefont  [1]{#1}%
\providecommand \bibfnamefont [1]{#1}%
\providecommand \citenamefont [1]{#1}%
\providecommand \href@noop [0]{\@secondoftwo}%
\providecommand \href [0]{\begingroup \@sanitize@url \@href}%
\providecommand \@href[1]{\@@startlink{#1}\@@href}%
\providecommand \@@href[1]{\endgroup#1\@@endlink}%
\providecommand \@sanitize@url [0]{\catcode `\\12\catcode `\$12\catcode
  `\&12\catcode `\#12\catcode `\^12\catcode `\_12\catcode `\%12\relax}%
\providecommand \@@startlink[1]{}%
\providecommand \@@endlink[0]{}%
\providecommand \url  [0]{\begingroup\@sanitize@url \@url }%
\providecommand \@url [1]{\endgroup\@href {#1}{\urlprefix }}%
\providecommand \urlprefix  [0]{URL }%
\providecommand \Eprint [0]{\href }%
\providecommand \doibase [0]{https://doi.org/}%
\providecommand \selectlanguage [0]{\@gobble}%
\providecommand \bibinfo  [0]{\@secondoftwo}%
\providecommand \bibfield  [0]{\@secondoftwo}%
\providecommand \translation [1]{[#1]}%
\providecommand \BibitemOpen [0]{}%
\providecommand \bibitemStop [0]{}%
\providecommand \bibitemNoStop [0]{.\EOS\space}%
\providecommand \EOS [0]{\spacefactor3000\relax}%
\providecommand \BibitemShut  [1]{\csname bibitem#1\endcsname}%
\let\auto@bib@innerbib\@empty
\bibitem [{\citenamefont {Assayag}\ \emph {et~al.}(1993)\citenamefont
  {Assayag}, \citenamefont {Vieu}, \citenamefont {Gierak}, \citenamefont
  {Sudraud},\ and\ \citenamefont {Corbin}}]{assayag1993}%
  \BibitemOpen
  \bibfield  {author} {\bibinfo {author} {\bibfnamefont {G.~B.}\ \bibnamefont
  {Assayag}}, \bibinfo {author} {\bibfnamefont {C.}~\bibnamefont {Vieu}},
  \bibinfo {author} {\bibfnamefont {J.}~\bibnamefont {Gierak}}, \bibinfo
  {author} {\bibfnamefont {P.}~\bibnamefont {Sudraud}},\ and\ \bibinfo {author}
  {\bibfnamefont {A.}~\bibnamefont {Corbin}},\ }\bibfield  {title} {\enquote
  {\bibinfo {title} {New characterization method of ion current-density profile
  based on damage distribution of ga+ focused-ion beam implantation in gaas},}\
  }\href@noop {} {\bibfield  {journal} {\bibinfo  {journal} {Journal of Vacuum
  Science \& Technology B: Microelectronics and Nanometer Structures
  Processing, Measurement, and Phenomena}\ }\textbf {\bibinfo {volume} {11}},\
  \bibinfo {pages} {2420--2426} (\bibinfo {year} {1993})}\BibitemShut {NoStop}%
\bibitem [{\citenamefont {Wang}\ and\ \citenamefont {Wang}(1996)}]{wang1996}%
  \BibitemOpen
  \bibfield  {author} {\bibinfo {author} {\bibfnamefont {J.~B.}\ \bibnamefont
  {Wang}}\ and\ \bibinfo {author} {\bibfnamefont {Y.~L.}\ \bibnamefont
  {Wang}},\ }\bibfield  {title} {\enquote {\bibinfo {title} {A novel procedure
  for measuring the absolute current density profile of a focused gallium-ion
  beam},}\ }\href@noop {} {\bibfield  {journal} {\bibinfo  {journal} {Applied
  physics letters}\ }\textbf {\bibinfo {volume} {69}},\ \bibinfo {pages}
  {2764--2766} (\bibinfo {year} {1996})}\BibitemShut {NoStop}%
\bibitem [{\citenamefont {Davis}\ \emph {et~al.}(1997)\citenamefont {Davis},
  \citenamefont {Bartsch}, \citenamefont {Olson}, \citenamefont {Rej},\ and\
  \citenamefont {Waganaar}}]{davis1997}%
  \BibitemOpen
  \bibfield  {author} {\bibinfo {author} {\bibfnamefont {H.~A.}\ \bibnamefont
  {Davis}}, \bibinfo {author} {\bibfnamefont {R.~R.}\ \bibnamefont {Bartsch}},
  \bibinfo {author} {\bibfnamefont {J.~C.}\ \bibnamefont {Olson}}, \bibinfo
  {author} {\bibfnamefont {D.~J.}\ \bibnamefont {Rej}},\ and\ \bibinfo {author}
  {\bibfnamefont {W.~J.}\ \bibnamefont {Waganaar}},\ }\bibfield  {title}
  {\enquote {\bibinfo {title} {Intense ion beam optimization and
  characterization with infrared imaging},}\ }\href@noop {} {\bibfield
  {journal} {\bibinfo  {journal} {Journal of Applied Physics}\ }\textbf
  {\bibinfo {volume} {82}},\ \bibinfo {pages} {3223--3231} (\bibinfo {year}
  {1997})}\BibitemShut {NoStop}%
\bibitem [{\citenamefont {Corr}\ and\ \citenamefont
  {Jacobs}(1992)}]{corr1992ion}%
  \BibitemOpen
  \bibfield  {author} {\bibinfo {author} {\bibfnamefont {D.}~\bibnamefont
  {Corr}}\ and\ \bibinfo {author} {\bibfnamefont {D.~C.}\ \bibnamefont
  {Jacobs}},\ }\bibfield  {title} {\enquote {\bibinfo {title} {An ion detector
  for imaging two-dimensional velocity distributions},}\ }\href@noop {}
  {\bibfield  {journal} {\bibinfo  {journal} {Review of scientific
  instruments}\ }\textbf {\bibinfo {volume} {63}},\ \bibinfo {pages}
  {1969--1972} (\bibinfo {year} {1992})}\BibitemShut {NoStop}%
\bibitem [{\citenamefont {Yu}\ \emph {et~al.}(2015)\citenamefont {Yu},
  \citenamefont {Shen}, \citenamefont {Qu}, \citenamefont {Liu}, \citenamefont
  {Zhong}, \citenamefont {Zhang}, \citenamefont {Zhang}, \citenamefont {Yan},
  \citenamefont {Zhang}, \citenamefont {Zhang} \emph {et~al.}}]{yu2015}%
  \BibitemOpen
  \bibfield  {author} {\bibinfo {author} {\bibfnamefont {X.}~\bibnamefont
  {Yu}}, \bibinfo {author} {\bibfnamefont {J.}~\bibnamefont {Shen}}, \bibinfo
  {author} {\bibfnamefont {M.}~\bibnamefont {Qu}}, \bibinfo {author}
  {\bibfnamefont {W.}~\bibnamefont {Liu}}, \bibinfo {author} {\bibfnamefont
  {H.}~\bibnamefont {Zhong}}, \bibinfo {author} {\bibfnamefont
  {J.}~\bibnamefont {Zhang}}, \bibinfo {author} {\bibfnamefont
  {Y.}~\bibnamefont {Zhang}}, \bibinfo {author} {\bibfnamefont
  {S.}~\bibnamefont {Yan}}, \bibinfo {author} {\bibfnamefont {G.}~\bibnamefont
  {Zhang}}, \bibinfo {author} {\bibfnamefont {X.}~\bibnamefont {Zhang}}, \emph
  {et~al.},\ }\bibfield  {title} {\enquote {\bibinfo {title} {Characterization
  and analysis of infrared imaging diagnostics for intense pulsed ion and
  electron beams},}\ }\href@noop {} {\bibfield  {journal} {\bibinfo  {journal}
  {Vacuum}\ }\textbf {\bibinfo {volume} {113}},\ \bibinfo {pages} {36--42}
  (\bibinfo {year} {2015})}\BibitemShut {NoStop}%
\bibitem [{\citenamefont {Pitters}\ \emph {et~al.}(2019)\citenamefont
  {Pitters}, \citenamefont {Breitenfeldt}, \citenamefont {Pinto}, \citenamefont
  {Pahl}, \citenamefont {Pikin}, \citenamefont {Shornikov},\ and\ \citenamefont
  {Wenander}}]{pitters2019}%
  \BibitemOpen
  \bibfield  {author} {\bibinfo {author} {\bibfnamefont {J.}~\bibnamefont
  {Pitters}}, \bibinfo {author} {\bibfnamefont {M.}~\bibnamefont
  {Breitenfeldt}}, \bibinfo {author} {\bibfnamefont {S.~D.}\ \bibnamefont
  {Pinto}}, \bibinfo {author} {\bibfnamefont {H.}~\bibnamefont {Pahl}},
  \bibinfo {author} {\bibfnamefont {A.}~\bibnamefont {Pikin}}, \bibinfo
  {author} {\bibfnamefont {A.}~\bibnamefont {Shornikov}},\ and\ \bibinfo
  {author} {\bibfnamefont {F.}~\bibnamefont {Wenander}},\ }\bibfield  {title}
  {\enquote {\bibinfo {title} {Pepperpot emittance measurements of ion beams
  from an electron beam ion source},}\ }\href@noop {} {\bibfield  {journal}
  {\bibinfo  {journal} {Nuclear Instruments and Methods in Physics Research
  Section A: Accelerators, Spectrometers, Detectors and Associated Equipment}\
  }\textbf {\bibinfo {volume} {922}},\ \bibinfo {pages} {28--35} (\bibinfo
  {year} {2019})}\BibitemShut {NoStop}%
\bibitem [{\citenamefont {Sosolik}\ \emph {et~al.}(2000)\citenamefont
  {Sosolik}, \citenamefont {Lavery}, \citenamefont {Dahl},\ and\ \citenamefont
  {Cooper}}]{sosolik2000}%
  \BibitemOpen
  \bibfield  {author} {\bibinfo {author} {\bibfnamefont {C.~E.}\ \bibnamefont
  {Sosolik}}, \bibinfo {author} {\bibfnamefont {A.~C.}\ \bibnamefont {Lavery}},
  \bibinfo {author} {\bibfnamefont {E.~B.}\ \bibnamefont {Dahl}},\ and\
  \bibinfo {author} {\bibfnamefont {B.~H.}\ \bibnamefont {Cooper}},\ }\bibfield
   {title} {\enquote {\bibinfo {title} {A technique for accurate measurements
  of ion beam current density using a faraday cup},}\ }\href@noop {} {\bibfield
   {journal} {\bibinfo  {journal} {Review of Scientific Instruments}\ }\textbf
  {\bibinfo {volume} {71}},\ \bibinfo {pages} {3326--3330} (\bibinfo {year}
  {2000})}\BibitemShut {NoStop}%
\bibitem [{Mac()}]{Macor}%
  \BibitemOpen
  \href@noop {} {}\bibinfo {note} {{MACOR is a registered trademark of Corning
  Incorporated, Corning, NY 14831}}\BibitemShut {NoStop}%
\bibitem [{\citenamefont {Dahl}(1998)}]{dahl1998}%
  \BibitemOpen
  \bibfield  {author} {\bibinfo {author} {\bibfnamefont {E.~B.}\ \bibnamefont
  {Dahl}},\ }\href@noop {} {\emph {\bibinfo {title} {Multi-state charge
  transfer dynamics and trapping of hyperthermal and low energy alkali ions}}}\
  (\bibinfo  {publisher} {Cornell University},\ \bibinfo {year}
  {1998})\BibitemShut {NoStop}%
\bibitem [{\citenamefont {Arfken}\ and\ \citenamefont
  {Weber}(2001)}]{Arfken2001}%
  \BibitemOpen
  \bibfield  {author} {\bibinfo {author} {\bibfnamefont {G.~B.}\ \bibnamefont
  {Arfken}}\ and\ \bibinfo {author} {\bibfnamefont {H.~J.}\ \bibnamefont
  {Weber}},\ }\href@noop {} {\emph {\bibinfo {title} {Mathematical Methods for
  Physicists}}},\ \bibinfo {edition} {5th}\ ed.\ (\bibinfo  {publisher}
  {Academic Press},\ \bibinfo {year} {2001})\ pp.\ \bibinfo {pages}
  {924--925}\BibitemShut {NoStop}%
\bibitem [{\citenamefont {Bracewell}(1965)}]{bracewell1965}%
  \BibitemOpen
  \bibfield  {author} {\bibinfo {author} {\bibfnamefont {R.}~\bibnamefont
  {Bracewell}},\ }\href@noop {} {\emph {\bibinfo {title} {The Fourier Transform
  and Its Applications}}}\ (\bibinfo  {publisher} {McGraw-Hill},\ \bibinfo
  {year} {1965})\ pp.\ \bibinfo {pages} {108--112}\BibitemShut {NoStop}%
\bibitem [{\citenamefont {Shyam}\ \emph {et~al.}(2015)\citenamefont {Shyam},
  \citenamefont {Kulkarni}, \citenamefont {Field}, \citenamefont {Srinadhu},
  \citenamefont {Cutshall}, \citenamefont {Harrell}, \citenamefont {Harriss},\
  and\ \citenamefont {Sosolik}}]{shyam2015}%
  \BibitemOpen
  \bibfield  {author} {\bibinfo {author} {\bibfnamefont {R.}~\bibnamefont
  {Shyam}}, \bibinfo {author} {\bibfnamefont {D.~D.}\ \bibnamefont {Kulkarni}},
  \bibinfo {author} {\bibfnamefont {D.~A.}\ \bibnamefont {Field}}, \bibinfo
  {author} {\bibfnamefont {E.~S.}\ \bibnamefont {Srinadhu}}, \bibinfo {author}
  {\bibfnamefont {D.~B.}\ \bibnamefont {Cutshall}}, \bibinfo {author}
  {\bibfnamefont {W.~R.}\ \bibnamefont {Harrell}}, \bibinfo {author}
  {\bibfnamefont {J.~E.}\ \bibnamefont {Harriss}},\ and\ \bibinfo {author}
  {\bibfnamefont {C.~E.}\ \bibnamefont {Sosolik}},\ }\bibfield  {title}
  {\enquote {\bibinfo {title} {First multicharged ion irradiation results from
  the cuebit facility at clemson university},}\ }in\ \href@noop {} {\emph
  {\bibinfo {booktitle} {AIP Conference Proceedings}}},\ Vol.\ \bibinfo
  {volume} {1640}\ (\bibinfo {organization} {American Institute of Physics},\
  \bibinfo {year} {2015})\ pp.\ \bibinfo {pages} {129--135}\BibitemShut
  {NoStop}%
\bibitem [{dec()}]{deconvolution_script}%
  \BibitemOpen
  \href@noop {} {}\bibinfo {note} {Deconvolution script available at
  \url{https://github.com/rhmatti/Current-Density-Deconvolution}}\BibitemShut
  {NoStop}%
\end{thebibliography}%


%

\end{document}